\documentclass[twocolumn,showpacs,amsmath,amssymb,prl,superscriptaddress,floatfix]{revtex4}

\usepackage{graphicx}
\usepackage{dcolumn}
\usepackage{bm}

\begin{document}

\title{Ultracold atoms in radio-frequency-dressed potentials beyond the rotating wave approximation}
\date{\today}

\author{S.~Hofferberth}
\email{hofferberth@atomchip.org}
\affiliation{%
Physikalisches Institut, Universit\"at Heidelberg, Philosophenweg 12, D-69120 Heidelberg, Germany}%
\author{B.~Fischer}
\affiliation{%
Physikalisches Institut, Universit\"at Heidelberg, Philosophenweg 12, D-69120 Heidelberg, Germany}%
\author{T.~Schumm}
\affiliation{Atominstitut der \"Osterreichischen Universit\"aten,
TU-Wien, Stadionallee 2, A-1020 Vienna, Austria }
\author{J.~Schmiedmayer}
\affiliation{Physikalisches Institut, Universit\"at Heidelberg, Philosophenweg 12, D-69120 Heidelberg, Germany}%
\affiliation{Atominstitut der \"Osterreichischen Universit\"aten,
TU-Wien, Stadionallee 2, A-1020 Vienna, Austria }
\author{I.~Lesanovsky}
\email{Igor.Lesanovsky@uibk.ac.at}
\affiliation{Physikalisches Institut, Universit\"at Heidelberg, Philosophenweg 12, D-69120 Heidelberg, Germany}%
\affiliation{Institut f\"{u}r Theoretische Physik, Universit\"{a}t Innsbruck, Technikerstr. 21a, A-6020 Innsbruck, Austria}

\date{\today}

\begin{abstract}\label{txt:abstract}
We study dressed Bose-Einstein condensates in an atom chip radio-frequency trap. We show that in this system sufficiently strong dressing can be achieved to cause the widely used rotating wave approximation (RWA) to break down. We present a full calculation of the atom - field coupling which shows that the non-RWA contributions quantitatively alter the shape of the emerging dressed adiabatic potentials. The non-RWA contributions furthermore lead to additional allowed transitions between dressed levels. We use RF spectroscopy of Bose-Einstein condensates trapped in the dressed state potentials to directly observe the transition from the RWA to the beyond-RWA regime.

\end{abstract}
\maketitle

\section{Introduction}
Using external oscillating fields in order to manipulate atoms is a well-established experimental technique. Quantum optics provides a description of such driven atoms in terms of dressed states \cite{Cohen1992}. These new eigenstates contain contributions from both the atom and the external (dressing) field. Consequently the atomic properties are altered with respect to the field free case. This gives rise to effects like the Autler-Townes \cite{Autler1955} splitting or electromagnetically induced transparency \cite{Harris1997, Lukin2003}. Moreover, the resulting atomic level shift can be utilized for manipulating the external degrees of freedom. In the optical regime the corresponding light shift is used to build traps by exploiting spatial intensity modulation of standing light waves \cite{Grimm2000}. Microwave adiabatic potentials have been proposed in \cite{Agosta1989}, and a detuned micro-wave has been used for trapping ultra cold Cs atoms \cite{Spreeuw1994}. In the radio-frequency (RF) domain the use of dressed Zeeman states for trapping neutral atoms was first proposed in \cite{Zobay2001} and has recently been successfully employed to build complex traps and interferometers \cite{Colombe2004,Schumm2005b,Hofferberth2006,Jo2007,White2006}. The dressing of Zeeman states has also been studied for neutrons \cite{Muskat1987}.

A common approximation which is generally used in the context of dressed states is the rotating wave approximation (RWA), where during the derivation of the equations of motion of the dressed system, rapidly oscillating terms are neglected \cite{Rabi1954,Cohen1992}. This approximation is valid if the frequency $\omega$ of the driving field is near-resonant with the coupled atomic transition $\omega_0$, i.e $\omega \approx \omega_0$, and the Rabi frequency of the driving field $\Omega_\text{R} \ll \omega$ is much smaller than its oscillation frequency \cite{Lembessis2005}.

In this paper, we show that in atom chip RF-traps both conditions for the validity of the RWA can be violated, i.e. locally detunings $\Delta = \omega -\omega_0$ and Rabi frequencies $\Omega_\text{R}$ become comparable to the driving frequency $\omega$. We use RF spectroscopy \cite{Martin1988} to investigate the dressed states and compare our data to the RWA calculation and to a numerically exact calculation in a second quantization picture. We find significant quantitative deviations from the RWA for the resulting adiabatic potentials, which is of relevance to recent experiments \cite{Schumm2005b,Hofferberth2006,Jo2007}.

\section{Radio-frequency dressed atomic hyperfine states}
In RF dressing of atoms the coupled states are the Zeeman-shifted magnetic sublevels of an atomic hyperfine state \cite{Rabi1954}. We denote the static magnetic field causing the Zeeman shift by $\mathbf{B}_\text{S}(\textbf{r})$. The atomic states are coupled by an oscillating magnetic field $\mathbf{B}_\text{RF}(\textbf{r}) e^{i \omega_\text{RF}t}$. In the dressed state formalism, the total Hamiltonian then reads
\begin{eqnarray}
  H&=&\mu |\mathbf{B}_\text{S}(\textbf{r})|F_\text{z}+ \hbar \omega_\text{RF} a^\dagger\nonumber
  a+\gamma\left[B_{\text{RF}\perp}(\textbf{r})a^\dagger+\text{h.c.}\right]F_\text{x}\\
  &&+\gamma\left[B_{\text{RF}\parallel}(\textbf{r})a^\dagger+\text{h.c.}\right]F_\text{z},\label{eq:Hamiltonian}
\end{eqnarray}
with $\mu=\mu_\text{B} g_\text{F}$ and $\gamma=\mu/(2\sqrt{\left<N\right>})$, where $\mu_\text{B}$ is Bohr\'{}s magneton and $g_\text{F}$ is the Land\'{e}-factor of the considered hyperfine state. $\left<N\right>$ is the average photon number of the dressing field, $\mathbf{F}$ is the operator of the total atomic spin, and $B_{\text{RF} \perp}(\textbf{r})$, $B_{\text{RF} \parallel}(\textbf{r})$ are the complex amplitudes of the RF field components perpendicular and parallel to the static field vector. $a^\dagger$ is the creation operator for quanta of the RF field.

The first term of the Hamiltonian describes the Zeeman shift of the atomic levels in the static field, while the second term accounts for the energy of the RF field. The coupling between the atomic levels and the dressing field is established by the third and the fourth term. The latter involves only components of $\mathbf{B}_\text{RF}(\textbf{r})$ that oscillate parallel to the static field and can be neglected if $|\mu\mathbf{B}_{\text{RF}\parallel}(\textbf{r})|\ll \hbar\omega$ \cite{Pegg1974}.

In order to study the non-RWA effects we diagonalize the full Hamiltonian (\ref{eq:Hamiltonian}) numerically for $F=2$ in the basis spanned by the bare states $\left\{\left|m_\text{F},\triangle N\right>\right\}$, where $m_\text{F}$ is the magnetic quantum number of the atomic level and $\triangle N = N - \left<N\right>$, with $N$ being the number of RF photons. The RF field is best described by a coherent state, i.e. a superposition of number states with a poissonian distribution around $\left<N\right>$. We are not interested in the change of the RF field during the coupling and assume $\left<N\right>$ to be large. Therefore we use $N+1 \approx N$ and only consider a small number of photon states centered around the mean photon number \cite{Allegrini1971}.

Under this assumption, that the RF field can be treated as a classical field, the dressed state formalism including the non-RWA terms is equivalent to the theory of Floquet states, as shown in \cite{Shirley1965}. In this semiclassical approach the quantization of the RF field is not explicitly included, but the Floquet states can be interpreted as quantum states containing a definite, very large, photon number. For our numerical calculation, we choose the dressed state picture, as it yields a more obvious connection to existing RWA treatments of the system \cite{Zobay2001,Lesanovsky2006,Lesanovsky2006b}.

It is convenient to group the bare states into manifolds $\left\{\left|m_\text{F},\kappa-\text{sgn}(g_\text{F})\,m_\text{F}\right>\right\},\,(m_\text{F}=-F,...,F)$
which are denoted by the number $\kappa$. How many manifolds are required for the calculation depends on the strength of the off-resonant contributions to the coupling term in Hamiltonian (\ref{eq:Hamiltonian}), which introduce a coupling between manifolds with $\left|\kappa-\kappa^\prime\right|=2$. In the numeric calculations we include 25 manifolds ($\triangle N =-12,...,12$) to avoid numerical artifacts.

Considering only a single $\kappa$-manifold of bare states in the diagonalization is equivalent to applying the RWA, in which case the resulting potentials take on the well known form \cite{Lesanovsky2006b}
\begin{eqnarray}
  V_\text{RWA}(\mathbf{r})=\tilde{m}_\text{F}\,\text{sgn}(\mu)
  \sqrt{\Delta^2(\mathbf{r})+\Omega^2(\mathbf{r})}\label{eq:RWA_potentials}
\end{eqnarray}
with the detuning $\Delta(\mathbf{r})=|\mu||\mathbf{B}_\text{S}(\textbf{r})|-\hbar\omega$ and the Rabi frequency $\Omega(\mathbf{r})=\frac{\mu}{2}|B_{\text{RF}\perp}(\textbf{r})|$. In this case, the resulting dressed states can be grouped in manifolds $\left| \tilde{m}_\text{F}, (\kappa) \right>$, where $\tilde{m}_\text{F}=-F,...,F$ is the effective magnetic quantum number of the dressed states. These manifolds can be characterized by a single $\kappa$, because in the RWA case each dressed state only contains contributions of bare states from one $\kappa$ manifold.

This is no longer true if the off-resonant terms become significant. Then each dressed state becomes a superposition of bare states from many manifolds. Still, for the coupling strengths considered here, it remains possible to identify groups of five dressed states each, with effective quantum numbers $\tilde{m_\text{F}}$. We will use this notation also to label the dressed states obtained from the full calculation.

\begin{figure}
\includegraphics[angle=0,width=\columnwidth]{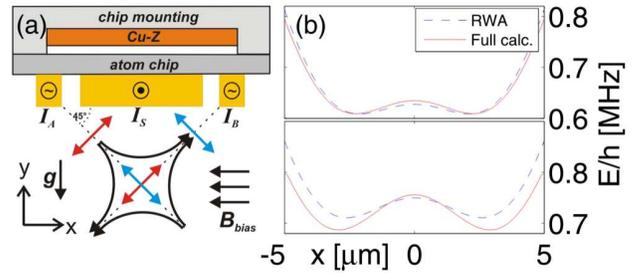}
\caption{(a) Schematic of the experimental setup. The central $100 \mu$m wide wire creates a static magnetic trap, which distance to the chip surface can be controlled by the wire current and the external bias field strength. For the experiments described here, the trap is positioned directly below the wire at a distance of $d=110 \,\mu$m from the chip surface. AC currents of $5...60$ mA are applied to the $10 \,\mu$m wide outer wires. The center-to-center distance between the central wire and the RF wires is $115 \,\mu$m. The amplitudes of the resulting dressing fields are $0.1...0.7$ G at the static trap center.
(b) Comparison of potentials obtained using the RWA (dashed lines) and the full calculation (solid line) for two different strengths of the dressing field ($I_\text{RF} = I_\text{A} = I_\text{B}=50$ and $60$ mA).}
\label{Fig:setup}
\end{figure}

\section{The experiment}
To experimentally realize the RF potentials we use a three wire atom chip setup as shown in Fig. \ref{Fig:setup}a. We prepare Bose-Einstein condensates (BECs) of $\sim10^4$ $^{87}$Rb atoms in the $\left|F=2, m_\text{F}=2\right>$ state in a standard Z-wire Ioffe-Pritchard micro trap formed by a DC current in the central wire and a homogeneous bias field \cite{Folman2002}. Our scheme of producing BEC in this trap is described in \cite{Wildermuth2004}.
To create the RF dressing field, AC currents with frequency $\omega_\text{RF}$ and amplitudes $I_\text{A}$ and $I_\text{B}$ are applied to two additional wires on the atom chip, one on each side of the Z-shaped wire (Fig. \ref{Fig:setup}a). The total dressing field reads $\left[\mathbf{B}_\text{A}(\textbf{r}) + \mathbf{B}_\text{B}(\textbf{r}) e^{i \delta}\right]e^{i \omega_\text{RF} t}$,
where $\delta$ is the phase shift between the two RF currents. This field configuration allows the realization of versatile RF potentials, for example a rotated double well or a ring shaped trap \cite{Lesanovsky2006,Hofferberth2006}.

In the experiments described here, the two RF currents are always equal $I_\text{A}=I_\text{B}=I_\text{RF}$, while the phase shift is set to $\delta=\pi$ and the frequency to $\omega_\text{RF} = 2\pi \times 600$ kHz. The parameters of the static trap are chosen such that $\omega_{\perp} = 2 \pi \times 3$ kHz and $\omega_{\parallel}=2 \pi \times 20$ Hz. The trap center is positioned $115$ $\mu$m away from the chip surface. The Larmor frequency of the trapped atoms at the trap minimum is $\omega_\text{L}=2 \pi \times 650$ kHz, so that the minimal detuning is $50$ kHz. The resulting dressed potential is a symmetric, horizontal double well, with the well separation and the barrier height being controlled by $I_\text{RF}$ (Fig. \ref{Fig:setup}b).

To calculate the beyond-RWA dressed RF potential of this configuration, we insert the static magnetic field $\mathbf{B}_\text{S}$ of the Z-wire trap and the complex amplitude $\mathbf{B}_\text{RF}(\textbf{r})=\mathbf{B}_\text{A}(\textbf{r}) + e^{i\delta} \mathbf{B}_\text{B}(\textbf{r})$ of the combined RF fields into the Hamiltonian (\ref{eq:Hamiltonian}) and perform the diagonalization. We observe that although qualitatively the potentials are similar, both the splitting distance and the barrier height are modified quantitatively (Fig. \ref{Fig:setup}b). The latter is changed by more than a factor two for the largest RF dressing fields we have studied. This is of importance for current RF double well experiments, since the tunneling rate between the wells depends exponentially on the potential barrier \cite{Smerzi1997}.

\begin{figure}
\includegraphics[angle=0,width=\columnwidth]{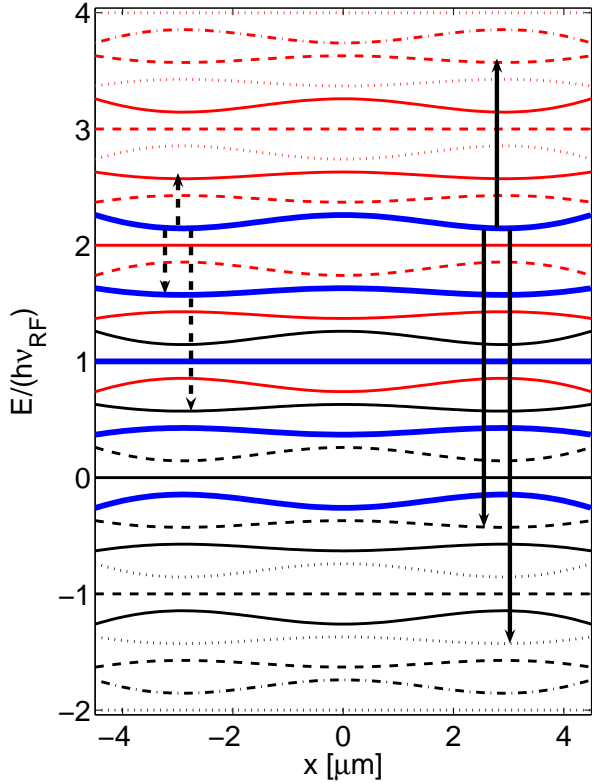}
\caption{Level scheme of the RF dressed states of the double well potential for $I_\text{RF}=60$ mA. Five levels associated with one $\kappa$-manifold are shown in bold lines. Different manifolds (indicated by color and linestyle) completely overlap for this coupling strength. The arrows mark the transitions induced by the "tickling" field, with the dashed arrows indicating those predicted by the RWA calculation. The full calculation leads to additional allowed transitions, indicated by the solid arrows (only the three energetically lowest are shown). Due to the symmetry of the potential, atoms are removed from both potential minima simultaneously.
}\label{Fig:spectroscopy}
\end{figure}

Fig. \ref{Fig:spectroscopy}a shows the dressed state level structure for an RF current of $I_\text{RF}=60$ mA. Five levels, which are associated with a single $\kappa$ and quantum numbers $\tilde{m}_\text{RF}=-2,...,2$ are highlighted. It can be seen that different manifolds completely overlap, making a clear separation impossible.

\section{RF spectroscopy of dressed states}
Experimentally, measuring the changes of the well separation and the potential barrier precisely is difficult. The well separation has to be inferred from interference patterns, which is complicated by atom-atom interaction during the expansion of the interfering BECs \cite{Schumm2005b}. The potential barrier could be measured by observing tunneling between the wells \cite{Albiez2005}. We investigate the modification of the RF dressed states due to the beyond-RWA contributions by performing a spectroscopic measurement \cite{Martin1988}. We measure the energy difference between dressed states by irradiating the dressed BEC with an additional weak RF "tickling" field $\mathbf{B}_\text{spec}(\textbf{r}) e^{i \omega_\text{spec} t}$ \cite{Happer1964}. If this field is resonant with the dressed state level spacing, transitions to untrapped states are induced. This results in trap loss, which is the signature for a resonance.

We calculate the allowed transitions using time-dependent perturbation theory, writing the operator of the spectroscopy field as $\mathbf{B}_\text{spec}(\textbf{r}) \cdot \mathbf{F}$.
This approach is valid only if the spectroscopy field does not deform the dressed states. In our experiments, we use $B_\text{spec} \approx 10^{-3} \times B_\text{RF}$. We verified experimentally that this treatment is justified by repeating our spectroscopy measurements with a doubled amplitude $B_\text{spec}$ and observed no measurable difference in the transition frequencies.

When calculating the transition matrix, we observe that non-vanishing elements only occur if $|\tilde{m}_\text{F}' - \tilde{m}_\text{F}| = 0,1$. This means, that although the dressed states are superpositions of all involved bare states, a selection rule similar to the case of RF-transitions between undressed states exists. This differs for example from spontaneous decay in optical dressing, where transitions between all dressed states can occur \cite{Mollow1969}.

If the RWA is applied to calculate the dressed states, only transitions with $|\kappa' - \kappa| = 0,1$ occur, resulting in a total of three allowed transitions (dashed arrows in Fig. \ref{Fig:spectroscopy}). This is due to the fact that the RWA dressed states only contain contributions from bare states of a single $\kappa$-manifold and that the spectroscopy operator does not act on the photon quantum number of the bare states. In contrast, the full numerical calculation predicts higher order transitions to occur. This is because bare states with different $\kappa$ contribute to each dressed state. This leads to a chain of allowed transition frequencies given by $\nu_\text{trans}=  n \omega_\text{RF}/(2 \pi) \pm \Omega$, where $n=0,1,2,...$ and $\Omega$ is the energy difference between dressed states within one $\kappa$-manifold (solid arrows in Fig. \ref{Fig:spectroscopy}) \cite{Haroche1970}. The calculated transition rates strongly depend on the amplitude of the dressing field. For increasing RF coupling, the higher order transitions become stronger. Additionally, the maximum transition rate is no longer located at $n=0$ but at higher $n$. For our parameters, at $I_\text{RF}=60$ mA the $n=2$ transition is strongest.

The experimental procedure for performing the spectroscopy is as follows: After transferring a BEC from the static trap into the RF potential, we switch on the weak spectroscopy field for a time $t_\text{spec}=100$ ms at frequency $\nu_\text{spec}$, while all other parameters are held constant. This field is generated by an AC current of $0.1$ mA applied to a macroscopic wire $1.2$ mm below the atom chip (Fig. \ref{Fig:setup}a). After the spectroscopy time we switch off all fields and measure the number of atoms by taking a time-of-flight absorption image of the released cloud. Between experiment cycles we vary $\nu_\text{spec}$ and search for frequencies at which we observe atom loss.
\begin{figure}
\includegraphics[angle=0,width=\columnwidth]{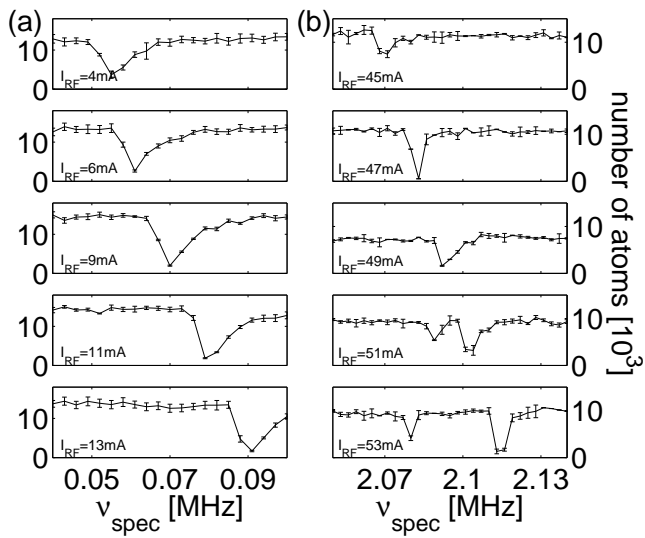}
\caption{(a) Frequency scans for increasing RF dressing current $I_\text{RF}$ in the range corresponding to the lowest lying transition. The signature of a transition is the loss of atoms from the trap. Each data point is the average of the observed atom number of $5-10$ experiments with identical parameters, reducing the signal noise due to shot-to-shot fluctuations of the atom number. The slightly asymmetric shape of the observed atom loss signal is due to the finite size and temperature of the probed BECs, which has to be considered in the determination of the resonance frequency. The position of the resonance shifts to higher frequency with the RF current, as expected.
(b)Similar scans of the frequency range corresponding to the crossing of the $3\times\omega_\text{RF}/(2 \pi) +\Omega$ and the $4\times\omega_\text{RF}/(2\pi)-\Omega$ non-RWA transitions at large RF currents. It can be seen that for the same spectroscopy time $t_\text{spec}=100$ ms and tickling field amplitude the lower transition rates of these resonances result in weaker atom loss. Specifically, the $4\times\omega_\text{RF}/(2\pi)-\Omega$ only becomes discernible for sufficiently large dressing fields.}\label{Fig:spec_scans}
\end{figure}

\section{Results}
Figure \ref{Fig:spec_scans} shows two sets of such scans at frequencies corresponding to the $n=0$ transition at low RF currents (fig. \ref{Fig:spec_scans}a) and to the crossing of the $3\times\omega_\text{RF}/(2 \pi) +\Omega$ and the $4\times\omega_\text{RF}/(2\pi)-\Omega$ non-RWA transitions (fig. \ref{Fig:spec_scans}b). For given spectroscopy time $t_\text{spec}$ and tickling field amplitude, the observed atom loss is directly proportional to the transition rates. It can be seen that the $4\times\omega_\text{RF}/(2\pi)-\Omega$ resonance only becomes strong enough to cause measurable loss of atoms for large RF currents. The observed transition rates are in good agreement with our numerical results.

In the exact determination of the resonance frequency and the comparison to calculations, the finite temperature and extension of the BEC and its gravitational sag in the potential have to be considered. Taking all errors into account we determine $\nu_\text{trans}$ with an accuracy of $\sim 1$ kHz in the case of the $n \leq 2$ resonances, which are used for the comparison with the RWA calculations in the following. To achieve the same accuracy for the weaker higher order transitions, the spectroscopy time $t_\text{spec}$ has to be increased.

The RF spectroscopy can also be used for evaporative cooling in the RF potentials, by sweeping $\nu_\text{trans}$ above a resonance, as experimentally demonstrated in \cite{Hofferberth2006}, and theoretically analyzed in\cite{Alzar2006}. We have efficiently cooled thermal ensembles to degeneracy using various RWA and non-RWA transitions. Additionally, we have selectively evaporated atoms from one side of an asymmetric double well.

\begin{figure}
\includegraphics[angle=0,width=\columnwidth]{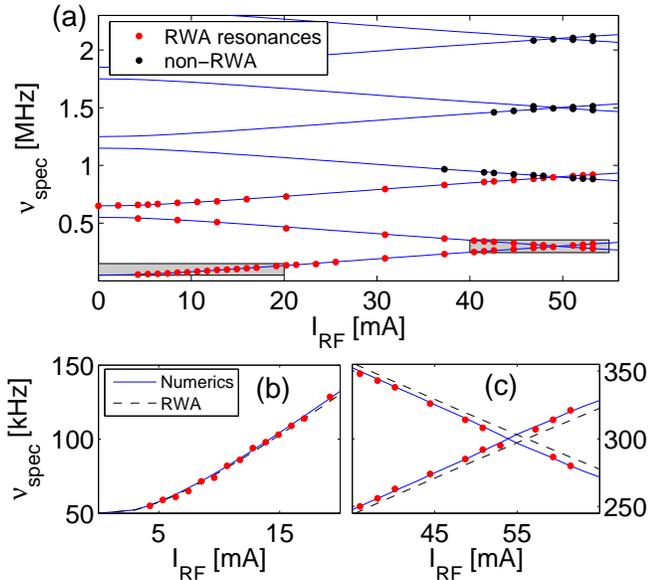}
\caption{(a) Observed resonances between $0...2.2$ MHz for $I_\text{RF}=0...60$ mA. The numerically calculated transition frequencies are shown as blue lines. It can be seen that for low $I_\text{RF}$ only three transitions are observable (red points). For higher RF currents additional resonances appear (blue squares). The error bars are smaller than the markers for all shown data points.
(b) Zoom ins into the two grey-shaded regions of plot (a). Both the numerically calculated transition frequencies (solid line) as well as those obtained from the RWA calculations are plotted (dashed line). For low RF amplitudes the RWA is in good agreement with the full calculation and both agree well with our measurements (left).  At higher RF amplitude the non-RWA terms lead to a shift of the resonances which can be measured by the RF spectroscopy (right).}
\label{Fig:Brf_scan}
\end{figure}

In Fig. \ref{Fig:Brf_scan} the result of a complete spectroscopy scan between $0$ and $2.2$ MHz for RF currents $I_\text{RF}=0...60$ mA is shown. In this scan the appearance of beyond-RWA effects for strong coupling fields can be seen. For low RF currents (small dressing field) we observe three transition frequencies, as predicted within the RWA. The non-RWA transitions are too weak to remove atoms from the trap in the spectroscopy time. The measured transition frequencies are in good agreement with those obtained by the RWA calculation, as can be seen in Fig. \ref{Fig:Brf_scan}b. For larger amplitudes of the dressing field, we observe additional transitions at higher frequencies, as predicted by the full calculation.

Furthermore, we observe a Bloch-Siegert shift of the transition frequencies from the RWA calculations (Fig. \ref{Fig:Brf_scan}c) \cite{Bloch1940, Wei1997}, which is also in excellent agreement with the full calculation. For the strongest coupling realized, this shift is on the order of $10$ kHz, which is one order of magnitude larger than the precision of our measurement. We verify that this effect is indeed a beyond-RWA effect and cannot be ascribed to an uncertainty of our experiment parameters. To this end we independently fit the RWA model to the data, using the field amplitudes as free parameters. This model fails to reproduce the shift of the resonance crossing while at the same time yielding good agreement with the observed resonances for small RF currents.
It has to be emphasized that we measure the shift on the energy difference between two dressed states, the absolute deviation between RWA and full calculation for individual dressed states is larger (Figure \ref{Fig:setup}b).

\section{Conclusion}
In conclusion, we have shown that RF dressed atoms in an atom chip trap are an ideally suited model system for studying effects beyond the RWA. It allows one to access both regimes in which the RWA breaks down, the realization of large coupling as well as (locally) large detuning compared to the resonance frequency. We found that in recent atom chip interference experiments \cite{Schumm2005b,Hofferberth2006,Jo2007} the RF coupling can get strong enough for beyond-RWA effects to become significant. A full calculation of the coupling becomes necessary for an accurate description of the adiabatic RF potentials. We experimentally verified the modifications beyond the RWA by carrying out RF spectroscopy on dressed BECs. We find that, beyond the transitions obtained in the RWA, additional higher order transitions occur, as predicted by a full calculation. The observed transition frequencies are in excellent agreement with the numerical results, while there is a clear deviation from the RWA. An improved accurate knowledge of the adiabatic potentials is very important in current experiments employing RF induced double well potentials, especially for the inferred tunnelling rates. Additionally the RF "tickling" field can be used for efficient evaporative cooling of RF dressed atoms, greatly enhancing the flexibility of RF potentials, allowing for example the study of coherence properties of independently created BECs \cite{Hofferberth2006}.

We thank A. Aspect, Ch. Westbrook, J. Dalibard, and J.H. Thywissen for helpful discussions and advise. We acknowledge financial support from the European Union, through the contracts MRTN-CT-2003-505032 (Atom Chips), Integrated Project FET/QIPC `SCALA'. I.L. acknowledges support from the European Community and its 6th Community Frame (program of Scholarships of Distinction `Marie Curie').


\begin{thebibliography}{32}
\expandafter\ifx\csname natexlab\endcsname\relax\def\natexlab#1{#1}\fi
\expandafter\ifx\csname bibnamefont\endcsname\relax
  \def\bibnamefont#1{#1}\fi
\expandafter\ifx\csname bibfnamefont\endcsname\relax
  \def\bibfnamefont#1{#1}\fi
\expandafter\ifx\csname citenamefont\endcsname\relax
  \def\citenamefont#1{#1}\fi
\expandafter\ifx\csname url\endcsname\relax
  \def\url#1{\texttt{#1}}\fi
\expandafter\ifx\csname urlprefix\endcsname\relax\def\urlprefix{URL }\fi
\providecommand{\bibinfo}[2]{#2}
\providecommand{\eprint}[2][]{\url{#2}}

\bibitem[{\citenamefont{Cohen-Tannoudji
  et~al.}(1992)\citenamefont{Cohen-Tannoudji, Dupont-Roc, and
  Grynberg}}]{Cohen1992}
\bibinfo{author}{\bibfnamefont{C.}~\bibnamefont{Cohen-Tannoudji}},
  \bibinfo{author}{\bibfnamefont{J.}~\bibnamefont{Dupont-Roc}},
  \bibnamefont{and} \bibinfo{author}{\bibfnamefont{G.}~\bibnamefont{Grynberg}},
  \emph{\bibinfo{title}{Atom-Photon Interactions}} (\bibinfo{publisher}{Wiley},
  \bibinfo{address}{New York}, \bibinfo{year}{1992}).

\bibitem[{\citenamefont{Autler and Townes}(1955)}]{Autler1955}
\bibinfo{author}{\bibfnamefont{S.~H.} \bibnamefont{Autler}} \bibnamefont{and}
  \bibinfo{author}{\bibfnamefont{C.~H.} \bibnamefont{Townes}},
  \bibinfo{journal}{Phys. Rev.} \textbf{\bibinfo{volume}{100}},
  \bibinfo{pages}{703} (\bibinfo{year}{1955}).

\bibitem[{\citenamefont{Harris}(1997)}]{Harris1997}
\bibinfo{author}{\bibfnamefont{S.~E.} \bibnamefont{Harris}},
  \bibinfo{journal}{Phys. Today} \textbf{\bibinfo{volume}{50}},
  \bibinfo{pages}{37} (\bibinfo{year}{1997}).

\bibitem[{\citenamefont{Lukin}(2003)}]{Lukin2003}
\bibinfo{author}{\bibfnamefont{M.~D.} \bibnamefont{Lukin}},
  \bibinfo{journal}{Rev. Mod. Phys} \textbf{\bibinfo{volume}{75}},
  \bibinfo{pages}{457} (\bibinfo{year}{2003}).

\bibitem[{\citenamefont{Grimm et~al.}(2000)\citenamefont{Grimm,
  Weidem{\"u}ller, and Ovchinnikov}}]{Grimm2000}
\bibinfo{author}{\bibfnamefont{R.}~\bibnamefont{Grimm}},
  \bibinfo{author}{\bibfnamefont{M.}~\bibnamefont{Weidem{\"u}ller}},
  \bibnamefont{and} \bibinfo{author}{\bibfnamefont{Y.~B.}
  \bibnamefont{Ovchinnikov}}, \bibinfo{journal}{Adv. At. Mol. Opt. Phys.}
  \textbf{\bibinfo{volume}{42}}, \bibinfo{pages}{95} (\bibinfo{year}{2000}).

\bibitem[{\citenamefont{Agosta et~al.}(1989)\citenamefont{Agosta, Silvera,
  Stoof, and Verhaar}}]{Agosta1989}
\bibinfo{author}{\bibfnamefont{C.~C.} \bibnamefont{Agosta}},
  \bibinfo{author}{\bibfnamefont{I.~F.} \bibnamefont{Silvera}},
  \bibinfo{author}{\bibfnamefont{H.~T.~C.} \bibnamefont{Stoof}},
  \bibnamefont{and} \bibinfo{author}{\bibfnamefont{B.~J.}
  \bibnamefont{Verhaar}}, \bibinfo{journal}{Phys. Rev. Lett.}
  \textbf{\bibinfo{volume}{62}}, \bibinfo{pages}{2361} (\bibinfo{year}{1989}).

\bibitem[{\citenamefont{Spreeuw et~al.}(1994)\citenamefont{Spreeuw, Gerz,
  Goldner, Phillips, Rolston, Westbrook, Reynolds, and Silvera}}]{Spreeuw1994}
\bibinfo{author}{\bibfnamefont{R.~J.~C.} \bibnamefont{Spreeuw}},
  \bibinfo{author}{\bibfnamefont{C.}~\bibnamefont{Gerz}},
  \bibinfo{author}{\bibfnamefont{L.~S.} \bibnamefont{Goldner}},
  \bibinfo{author}{\bibfnamefont{W.~D.} \bibnamefont{Phillips}},
  \bibinfo{author}{\bibfnamefont{S.~L.} \bibnamefont{Rolston}},
  \bibinfo{author}{\bibfnamefont{C.~I.} \bibnamefont{Westbrook}},
  \bibinfo{author}{\bibfnamefont{M.~W.} \bibnamefont{Reynolds}},
  \bibnamefont{and} \bibinfo{author}{\bibfnamefont{I.~F.}
  \bibnamefont{Silvera}}, \bibinfo{journal}{Phys. Rev. Lett.}
  \textbf{\bibinfo{volume}{72}}, \bibinfo{pages}{3162} (\bibinfo{year}{1994}).

\bibitem[{\citenamefont{Zobay and Garraway}(2001)}]{Zobay2001}
\bibinfo{author}{\bibfnamefont{O.}~\bibnamefont{Zobay}} \bibnamefont{and}
  \bibinfo{author}{\bibfnamefont{B.~M.} \bibnamefont{Garraway}},
  \bibinfo{journal}{Phys. Rev. Lett.} \textbf{\bibinfo{volume}{86}},
  \bibinfo{pages}{1195} (\bibinfo{year}{2001}).

\bibitem[{\citenamefont{Colombe et~al.}(2004)\citenamefont{Colombe, Knyazchyan,
  Morizot, Mercier, Lorent, and Perrin}}]{Colombe2004}
\bibinfo{author}{\bibfnamefont{Y.}~\bibnamefont{Colombe}},
  \bibinfo{author}{\bibfnamefont{E.}~\bibnamefont{Knyazchyan}},
  \bibinfo{author}{\bibfnamefont{O.}~\bibnamefont{Morizot}},
  \bibinfo{author}{\bibfnamefont{B.}~\bibnamefont{Mercier}},
  \bibinfo{author}{\bibfnamefont{V.}~\bibnamefont{Lorent}}, \bibnamefont{and}
  \bibinfo{author}{\bibfnamefont{H.}~\bibnamefont{Perrin}},
  \bibinfo{journal}{Europhys. Lett.} \textbf{\bibinfo{volume}{67}},
  \bibinfo{pages}{593} (\bibinfo{year}{2004}).

\bibitem[{\citenamefont{Schumm et~al.}(2005)\citenamefont{Schumm, Hofferberth,
  Andersson, Wildermuth, Groth, Bar-Joseph, Schmiedmayer, and
  Kr{\"u}ger}}]{Schumm2005b}
\bibinfo{author}{\bibfnamefont{T.}~\bibnamefont{Schumm}},
  \bibinfo{author}{\bibfnamefont{S.}~\bibnamefont{Hofferberth}},
  \bibinfo{author}{\bibfnamefont{L.~M.} \bibnamefont{Andersson}},
  \bibinfo{author}{\bibfnamefont{S.}~\bibnamefont{Wildermuth}},
  \bibinfo{author}{\bibfnamefont{S.}~\bibnamefont{Groth}},
  \bibinfo{author}{\bibfnamefont{I.}~\bibnamefont{Bar-Joseph}},
  \bibinfo{author}{\bibfnamefont{J.}~\bibnamefont{Schmiedmayer}},
  \bibnamefont{and}
  \bibinfo{author}{\bibfnamefont{P.}~\bibnamefont{Kr{\"u}ger}},
  \bibinfo{journal}{Nature Phys.} \textbf{\bibinfo{volume}{1}},
  \bibinfo{pages}{57} (\bibinfo{year}{2005}).

\bibitem[{\citenamefont{Hofferberth et~al.}(2006)\citenamefont{Hofferberth,
  Lesanovsky, Fischer, Verdu, and Schmiedmayer}}]{Hofferberth2006}
\bibinfo{author}{\bibfnamefont{S.}~\bibnamefont{Hofferberth}},
  \bibinfo{author}{\bibfnamefont{I.}~\bibnamefont{Lesanovsky}},
  \bibinfo{author}{\bibfnamefont{B.}~\bibnamefont{Fischer}},
  \bibinfo{author}{\bibfnamefont{J.}~\bibnamefont{Verdu}}, \bibnamefont{and}
  \bibinfo{author}{\bibfnamefont{J.}~\bibnamefont{Schmiedmayer}},
  \bibinfo{journal}{Nature Phys.} \textbf{\bibinfo{volume}{2}},
  \bibinfo{pages}{710} (\bibinfo{year}{2006}).

\bibitem[{\citenamefont{Jo et~al.}(2007)\citenamefont{Jo, Shin, Will, Pasquini,
  Saba, Ketterle, Pritchard, Vengalattore, and Prentiss}}]{Jo2007}
\bibinfo{author}{\bibfnamefont{G.-B.} \bibnamefont{Jo}},
  \bibinfo{author}{\bibfnamefont{Y.}~\bibnamefont{Shin}},
  \bibinfo{author}{\bibfnamefont{S.}~\bibnamefont{Will}},
  \bibinfo{author}{\bibfnamefont{T.~A.} \bibnamefont{Pasquini}},
  \bibinfo{author}{\bibfnamefont{M.}~\bibnamefont{Saba}},
  \bibinfo{author}{\bibfnamefont{W.}~\bibnamefont{Ketterle}},
  \bibinfo{author}{\bibfnamefont{D.~E.} \bibnamefont{Pritchard}},
  \bibinfo{author}{\bibfnamefont{M.}~\bibnamefont{Vengalattore}},
  \bibnamefont{and} \bibinfo{author}{\bibfnamefont{M.}~\bibnamefont{Prentiss}},
  \bibinfo{journal}{Phys. Rev. Lett.} \textbf{\bibinfo{volume}{98}}
  (\bibinfo{year}{2007}).

\bibitem[{\citenamefont{White et~al.}(2006)\citenamefont{White, Gao, Pasienski,
  and DeMarco}}]{White2006}
\bibinfo{author}{\bibfnamefont{M.}~\bibnamefont{White}},
  \bibinfo{author}{\bibfnamefont{H.}~\bibnamefont{Gao}},
  \bibinfo{author}{\bibfnamefont{M.}~\bibnamefont{Pasienski}},
  \bibnamefont{and} \bibinfo{author}{\bibfnamefont{B.}~\bibnamefont{DeMarco}},
  \bibinfo{journal}{Phys. Rev. A} \textbf{\bibinfo{volume}{74}}
  (\bibinfo{year}{2006}), \bibinfo{note}{023616}.

\bibitem[{\citenamefont{Muskat et~al.}(1987)\citenamefont{Muskat, Dubbers, and
  Sch{\"a}rpf}}]{Muskat1987}
\bibinfo{author}{\bibfnamefont{E.}~\bibnamefont{Muskat}},
  \bibinfo{author}{\bibfnamefont{D.}~\bibnamefont{Dubbers}}, \bibnamefont{and}
  \bibinfo{author}{\bibfnamefont{O.}~\bibnamefont{Sch{\"a}rpf}},
  \bibinfo{journal}{Phys. Rev. Lett.} \textbf{\bibinfo{volume}{58}},
  \bibinfo{pages}{2047} (\bibinfo{year}{1987}).

\bibitem[{\citenamefont{Rabi et~al.}(1954)\citenamefont{Rabi, Ramsey, and
  Schwinger}}]{Rabi1954}
\bibinfo{author}{\bibfnamefont{I.~I.} \bibnamefont{Rabi}},
  \bibinfo{author}{\bibfnamefont{N.~F.} \bibnamefont{Ramsey}},
  \bibnamefont{and}
  \bibinfo{author}{\bibfnamefont{J.}~\bibnamefont{Schwinger}},
  \bibinfo{journal}{Rev. Mod. Phys.} \textbf{\bibinfo{volume}{26}},
  \bibinfo{pages}{167} (\bibinfo{year}{1954}).

\bibitem[{\citenamefont{Lembessis and Ellinas}(2005)}]{Lembessis2005}
\bibinfo{author}{\bibfnamefont{V.~E.} \bibnamefont{Lembessis}}
  \bibnamefont{and} \bibinfo{author}{\bibfnamefont{D.}~\bibnamefont{Ellinas}},
  \bibinfo{journal}{J. Opt. B: Quantum Semiclass. Opt.}
  \textbf{\bibinfo{volume}{7}}, \bibinfo{pages}{319–322}
  (\bibinfo{year}{2005}).

\bibitem[{\citenamefont{Martin et~al.}(1988)\citenamefont{Martin, Helmerson,
  Bagnato, Lafyatis, and Pritchard}}]{Martin1988}
\bibinfo{author}{\bibfnamefont{A.~G.} \bibnamefont{Martin}},
  \bibinfo{author}{\bibfnamefont{K.}~\bibnamefont{Helmerson}},
  \bibinfo{author}{\bibfnamefont{V.~S.} \bibnamefont{Bagnato}},
  \bibinfo{author}{\bibfnamefont{G.~P.} \bibnamefont{Lafyatis}},
  \bibnamefont{and} \bibinfo{author}{\bibfnamefont{D.~E.}
  \bibnamefont{Pritchard}}, \bibinfo{journal}{Phys. Rev. Lett.}
  \textbf{\bibinfo{volume}{61}}, \bibinfo{pages}{2431} (\bibinfo{year}{1988}).

\bibitem[{\citenamefont{Pegg}(1974)}]{Pegg1974}
\bibinfo{author}{\bibfnamefont{D.~T.} \bibnamefont{Pegg}}, \bibinfo{journal}{J.
  Phys. B: At. Mol. Opt. Phys.} \textbf{\bibinfo{volume}{6}},
  \bibinfo{pages}{241 } (\bibinfo{year}{1974}).

\bibitem[{\citenamefont{Allegrini and Arimondo}(1971)}]{Allegrini1971}
\bibinfo{author}{\bibfnamefont{M.}~\bibnamefont{Allegrini}} \bibnamefont{and}
  \bibinfo{author}{\bibfnamefont{E.}~\bibnamefont{Arimondo}},
  \bibinfo{journal}{J. Phys. B: At. Mol. Phys.} \textbf{\bibinfo{volume}{4}},
  \bibinfo{pages}{1008} (\bibinfo{year}{1971}).

\bibitem[{\citenamefont{Shirley}(1965)}]{Shirley1965}
\bibinfo{author}{\bibfnamefont{J.~H.} \bibnamefont{Shirley}},
  \bibinfo{journal}{Phys. Rev.} \textbf{\bibinfo{volume}{138}},
  \bibinfo{pages}{B979} (\bibinfo{year}{1965}).

\bibitem[{\citenamefont{Lesanovsky
  et~al.}(2006{\natexlab{a}})\citenamefont{Lesanovsky, Schumm, Hofferberth,
  Andersson, Kr{\"u}ger, and Schmiedmayer}}]{Lesanovsky2006}
\bibinfo{author}{\bibfnamefont{I.}~\bibnamefont{Lesanovsky}},
  \bibinfo{author}{\bibfnamefont{T.}~\bibnamefont{Schumm}},
  \bibinfo{author}{\bibfnamefont{S.}~\bibnamefont{Hofferberth}},
  \bibinfo{author}{\bibfnamefont{L.~M.} \bibnamefont{Andersson}},
  \bibinfo{author}{\bibfnamefont{P.}~\bibnamefont{Kr{\"u}ger}},
  \bibnamefont{and}
  \bibinfo{author}{\bibfnamefont{J.}~\bibnamefont{Schmiedmayer}},
  \bibinfo{journal}{Phys. Rev. A} \textbf{\bibinfo{volume}{73}},
  \bibinfo{pages}{033619} (\bibinfo{year}{2006}{\natexlab{a}}).

\bibitem[{\citenamefont{Lesanovsky
  et~al.}(2006{\natexlab{b}})\citenamefont{Lesanovsky, Hofferberth,
  Schmiedmayer, and Schmelcher}}]{Lesanovsky2006b}
\bibinfo{author}{\bibfnamefont{I.}~\bibnamefont{Lesanovsky}},
  \bibinfo{author}{\bibfnamefont{S.}~\bibnamefont{Hofferberth}},
  \bibinfo{author}{\bibfnamefont{J.}~\bibnamefont{Schmiedmayer}},
  \bibnamefont{and}
  \bibinfo{author}{\bibfnamefont{P.}~\bibnamefont{Schmelcher}},
  \bibinfo{journal}{Phys. Rev. A.} \textbf{\bibinfo{volume}{74}},
  \bibinfo{pages}{033619} (\bibinfo{year}{2006}{\natexlab{b}}).

\bibitem[{\citenamefont{Folman et~al.}(2002)\citenamefont{Folman, Kr{\"u}ger,
  Schmiedmayer, Denschlag, and Henkel}}]{Folman2002}
\bibinfo{author}{\bibfnamefont{R.}~\bibnamefont{Folman}},
  \bibinfo{author}{\bibfnamefont{P.}~\bibnamefont{Kr{\"u}ger}},
  \bibinfo{author}{\bibfnamefont{J.}~\bibnamefont{Schmiedmayer}},
  \bibinfo{author}{\bibfnamefont{J.}~\bibnamefont{Denschlag}},
  \bibnamefont{and} \bibinfo{author}{\bibfnamefont{C.}~\bibnamefont{Henkel}},
  \bibinfo{journal}{Adv. At. Mol. Opt. Phys.} \textbf{\bibinfo{volume}{48}},
  \bibinfo{pages}{263} (\bibinfo{year}{2002}).

\bibitem[{\citenamefont{Wildermuth et~al.}(2004)\citenamefont{Wildermuth,
  Kr{\"u}ger, Becker, Brajdic, Haupt, Kasper, Folman, and
  Schmiedmayer}}]{Wildermuth2004}
\bibinfo{author}{\bibfnamefont{S.}~\bibnamefont{Wildermuth}},
  \bibinfo{author}{\bibfnamefont{P.}~\bibnamefont{Kr{\"u}ger}},
  \bibinfo{author}{\bibfnamefont{C.}~\bibnamefont{Becker}},
  \bibinfo{author}{\bibfnamefont{M.}~\bibnamefont{Brajdic}},
  \bibinfo{author}{\bibfnamefont{S.}~\bibnamefont{Haupt}},
  \bibinfo{author}{\bibfnamefont{A.}~\bibnamefont{Kasper}},
  \bibinfo{author}{\bibfnamefont{R.}~\bibnamefont{Folman}}, \bibnamefont{and}
  \bibinfo{author}{\bibfnamefont{J.}~\bibnamefont{Schmiedmayer}},
  \bibinfo{journal}{Phys. Rev. A} \textbf{\bibinfo{volume}{69}},
  \bibinfo{pages}{030901} (\bibinfo{year}{2004}).

\bibitem[{\citenamefont{Smerzi et~al.}(1997)\citenamefont{Smerzi, Fantoni,
  Giovanazzi, and Shenoy}}]{Smerzi1997}
\bibinfo{author}{\bibfnamefont{A.}~\bibnamefont{Smerzi}},
  \bibinfo{author}{\bibfnamefont{S.}~\bibnamefont{Fantoni}},
  \bibinfo{author}{\bibfnamefont{S.}~\bibnamefont{Giovanazzi}},
  \bibnamefont{and} \bibinfo{author}{\bibfnamefont{S.~R.}
  \bibnamefont{Shenoy}}, \bibinfo{journal}{Phys. Rev. Lett.}
  \textbf{\bibinfo{volume}{79}}, \bibinfo{pages}{4950} (\bibinfo{year}{1997}).

\bibitem[{\citenamefont{Albiez et~al.}(2005)\citenamefont{Albiez, Gati,
  F{\"o}lling, Hunsmann, Cristiani, and Oberthaler}}]{Albiez2005}
\bibinfo{author}{\bibfnamefont{M.}~\bibnamefont{Albiez}},
  \bibinfo{author}{\bibfnamefont{R.}~\bibnamefont{Gati}},
  \bibinfo{author}{\bibfnamefont{J.}~\bibnamefont{F{\"o}lling}},
  \bibinfo{author}{\bibfnamefont{S.}~\bibnamefont{Hunsmann}},
  \bibinfo{author}{\bibfnamefont{M.}~\bibnamefont{Cristiani}},
  \bibnamefont{and} \bibinfo{author}{\bibfnamefont{M.~K.}
  \bibnamefont{Oberthaler}}, \bibinfo{journal}{Phys. Rev. Lett.}
  \textbf{\bibinfo{volume}{95}}, \bibinfo{pages}{010402}
  (\bibinfo{year}{2005}).

\bibitem[{\citenamefont{Happer}(1964)}]{Happer1964}
\bibinfo{author}{\bibfnamefont{W.}~\bibnamefont{Happer}},
  \bibinfo{journal}{Phys. Rev.} \textbf{\bibinfo{volume}{136}},
  \bibinfo{pages}{A35} (\bibinfo{year}{1964}).

\bibitem[{\citenamefont{Mollow}(1969)}]{Mollow1969}
\bibinfo{author}{\bibfnamefont{B.~R.} \bibnamefont{Mollow}},
  \bibinfo{journal}{Phys. Rev.} \textbf{\bibinfo{volume}{188}},
  \bibinfo{pages}{1969} (\bibinfo{year}{1969}).

\bibitem[{\citenamefont{Haroche et~al.}(1970)\citenamefont{Haroche,
  Cohen-Tannoudji, Audoin, and Schermann}}]{Haroche1970}
\bibinfo{author}{\bibfnamefont{S.}~\bibnamefont{Haroche}},
  \bibinfo{author}{\bibfnamefont{C.}~\bibnamefont{Cohen-Tannoudji}},
  \bibinfo{author}{\bibfnamefont{C.}~\bibnamefont{Audoin}}, \bibnamefont{and}
  \bibinfo{author}{\bibfnamefont{J.~P.} \bibnamefont{Schermann}},
  \bibinfo{journal}{Phys. Rev. Lett.} \textbf{\bibinfo{volume}{24}},
  \bibinfo{pages}{861 } (\bibinfo{year}{1970}).

\bibitem[{\citenamefont{Alzar et~al.}(2006)\citenamefont{Alzar, Perrin,
  Garraway, and Lorent}}]{Alzar2006}
\bibinfo{author}{\bibfnamefont{C.~L.~G.} \bibnamefont{Alzar}},
  \bibinfo{author}{\bibfnamefont{H.}~\bibnamefont{Perrin}},
  \bibinfo{author}{\bibfnamefont{B.~M.} \bibnamefont{Garraway}},
  \bibnamefont{and} \bibinfo{author}{\bibfnamefont{V.}~\bibnamefont{Lorent}},
  \bibinfo{journal}{Phys. Rev. A} \textbf{\bibinfo{volume}{74}},
  \bibinfo{pages}{053413} (\bibinfo{year}{2006}).

\bibitem[{\citenamefont{Bloch and Siegert}(1940)}]{Bloch1940}
\bibinfo{author}{\bibfnamefont{F.}~\bibnamefont{Bloch}} \bibnamefont{and}
  \bibinfo{author}{\bibfnamefont{A.}~\bibnamefont{Siegert}},
  \bibinfo{journal}{Phys. Rev.} \textbf{\bibinfo{volume}{57}},
  \bibinfo{pages}{522} (\bibinfo{year}{1940}).

\bibitem[{\citenamefont{Wei et~al.}(1997)\citenamefont{Wei, Windsor, and
  Manson}}]{Wei1997}
\bibinfo{author}{\bibfnamefont{C.}~\bibnamefont{Wei}},
  \bibinfo{author}{\bibfnamefont{A.~S.~M.} \bibnamefont{Windsor}},
  \bibnamefont{and} \bibinfo{author}{\bibfnamefont{N.~B.}
  \bibnamefont{Manson}}, \bibinfo{journal}{J. Phys. B: At. Mol. Opt. Phys.}
  \textbf{\bibinfo{volume}{30}}, \bibinfo{pages}{4877 – 4888}
  (\bibinfo{year}{1997}).

\end{thebibliography}
\end{document}